# Adaptive Agent-Based SCADA System

Hosny Abbas[1], Samir Shaheen[2], and Mohammed Amin[3]

*Abstract—* Modern supervisory control and data acquisition (SCADA) systems comprise variety of industrial equipment such as physical control processes, logical control systems, communication networks, computers, and communication protocols. They are concerned with control and supervision of production control processes. Modern SCADA networks contain highly distributed information, control, and location. Moreover, they contain large number of heterogeneous components situated in highly changing and uncertain environments. As a result, engineering modern SCADA is a challenging issue and conventional engineering approaches are no longer suitable for them because of their increasing complexity and highly distribution. In this research, Multi-Agent Systems (MAS) are used to enable building adaptive agent-based SCADA system by modeling system components as agents in the micro level and as organizations or societies of agents in the macro level. A prototype has been implemented and evaluated within a simulation environment for demonstrating the adaptive behavior of the system-to-be, which results in continuous improvement of system performance.

*Keywords*—Adaptive SCADA, agent-based SCADA, complexity, multi-agent systems

## I. Introduction

MODERN SCADA networks are considered as subclass of Cyber-Physical Systems (CPS) [1], which are defined as the integrations of computation, networking, and physical processes. In CPS, embedded computers and networks monitor and control the physical processes, with feedback loops where physical processes affect computations and vice versa. The main concern of CPS is to bridge the cyber-world of computing and communications with the physical world. CPS are themselves subclass of Complex Adaptive Systems (CAS) [2] which are fluidly changing collections of distributed interacting components that react to both their environments and to one another. The main characteristics of these systems are highly distribution, continuous evolution, emergence, and complexity. Accordingly, modern SCADA systems may inherit from their super classes their complexity and challenges and require new engineering paradigms and approaches able to handle their challenges and characteristics not only in design time but also in run-time. SCADA systems need to be monitored, coordinated, controlled and integrated by a computing and communication infrastructure to achieve their required goals such as increasing productivity, reducing costs and sharing information in real-time across many industrial and enterprise systems. The way to achieve that is through building adaptive large-scale industrial networks integrated with recent information technologies, standard software systems and global communication networks such as the Internet. The benefits of this trend are: (1) it can lead to reduced development, operational, and maintenance costs, (2) providing executive management with real-time information that can be used to support planning, supervision, and decision making. Unfortunately, there are many challenges facing engineers and developers of this type of systems, in general these challenges can be described as quality attributes or non-functional attributes such as complexity, scalability, flexibility, Adaptivity, highly changing and uncertain working environments, reliability, security, etc. Complexity itself is not a quality attribute but the ability to handle complexity is. As stated in [3] complexity of the near future and even present applications can be characterized as a combination of aspects such as the great number of components taking part in the applications, the knowledge and control have to be distributed, the presence of non-linear processes in the system, the fact that the system is more and more often open, its environment dynamic and the interactions unpredictable. Scalability is another important quality attribute for modern SCADA systems because these systems required being open systems and as a result their size tends to increase as time goes, the increase in complexity is not only related to the number of system components but also to the amount of exchange data, the system should continue performing its function with good performance and quality as its size increases. Flexibility refers to systems that can adapt when changes occur in working environments. The working environments of modern SCADA systems are dynamic and changes continuously in unpredictable and uncertain manner. If the system is flexible enough, it will be able to adapt environment changes in run-time without the need to administrator intervention or at least with less intervention. These and more other challenges of modern SCADA are identified and described in details in [4].

Modern SCADA systems are required to possess simultaneously many quality attributes to survive and to continue working with higher quality of service (QoS) because they can be used to supervise and control critical nations' infrastructures and utilities such as power grids, water transportation, and oil and gas utilities, etc. Conventional engineering approaches and tools such as development methodologies, architectural styles, modeling techniques have limited capabilities to deal simultaneously with many quality attributes and require some important initial knowledge about the exact purposes of the system and every interaction to

Hosny Abbas[1] is a PhD student at Electrical Engineering Department, Assiut University, Egypt.
Samir Shaheen [2], is a professor of Computer Engineering, Cairo University, Egypt .
Mohammed Amin[3] is a professor of Control Engineering, Assiut University, Egypt.

which it may be confronted in the future have to be known in design time [5]. The feasible solution to the problem is through building adaptive systems capable of efficiently adapting to failures, component replacements and changes in the environment with less human intervention or centralized management. If a system is adaptive, it implicitly means that it's flexible able to adapt dynamic environment changes, scalable able to mange the increase in size, and also able to handle the evolution of its complexity. Nature is the most representative system of adaptivity that is why industrial engineers and academic researchers directed towards naturally-inspired models and techniques because nature is successful in this trend. Examples of naturally-inspired models are the models inspired from human social life, biology, physics, etc, and examples of naturally inspired techniques are genetic algorithms, evolutionary algorithms, self-organization, emergence, etc. From the other hand, realizing naturally-inspired models and techniques requires novel engineering styles and paradigms. One of the novel engineering architectural styles is MAS, which emerged as a scientific area from the previous research efforts in distributed artificial intelligence started in the early eighties. MAS are now seen as a major trend in research and development, they are mainly related to artificial intelligence and distributed computing techniques and are considered as the most representative among artificial systems dealing with complexity and distribution [3][6]. They have attracted great attention in many application domains where difficult and inherently distributed problems have to be tackled [7]. MAS provide an approach to solve a software problem by decomposing the system into a number of autonomous entities embedded in an environment in order to achieve the functional and quality requirements of the system [8].

Autonomy and Adaptivity are the main concepts behind MAS. Building adaptive MAS able to handle openness, complexity, and highly distribution of modern real world applications has recently attracted great attention. Adaptive MAS are designed to be capable to adapt themselves to unforeseen situations in an autonomous manner. They can be realized by enabling the system to dynamically reorganize (change its structure) to adapt dynamic environment changes [9]. Ferber et al. [10] pointed out that the classical agent-centered MAS (ACMAS) have many drawbacks and are no longer suitable to build complex software systems, and they stated that the solution can be achieved by using organization-centered MAS (OCMAS) in which higher order abstractions such as groups, organizations, and societies of agents should be considered as first order citizens within MAS. Designing and engineering OCMAS is usually done by using what is called organizational model [11]. The motivation to design organizational models is that in open environments, agents must be able to adapt towards the most appropriate organizations according to the working environment conditions and its dynamic unpredictable changes. As a result, organizational models should guarantee the ability of organizations to dynamically reorganize as a response to dynamic environment changes. In this paper, a dynamic organizational model for the analysis and design of OCMAS called NOSHAPE is proposed and used to build adaptive agent-based global SCADA. The NOSHAPE organizational model realizes dynamic reorganization within MAS through the overlapping of higher order entities (i.e. organizations of agents, worlds of organizations, and universes of worlds).

Often SCADA systems adopt the client-server architecture which is suitable for the domain specific activities. Therefore, this architecture will be used in the proposed system but will be augmented by adaptation architecture enables the system-to-be to adapt environments changes and operators new requirements and preferences. The adaptation architecture is based on the proposed MAS organizational model. The remaining of this paper is organized as follows: Section 2 provides a short background of MAS organization. Section 3 describes the proposed NOSHAPE organizational model. Section 4 provides the proposed adaptive agent-based SCADA. And section 5 concludes the paper and highlights future intentions.

## II. BACKGROUND

This section provides a short background of MAS organization and the next section presents the proposed organizational model. The domain specific activities will be presented later. Shehory [12] defined MAS organization as the way in which multiple agents are organized to form a MAS including the relationships and interactions among the agents and specific roles of agents within the organization. Jennings and Wooldridge [13] stated that considering MAS with no real structure is not suitable for handling current software systems complexity, and higher level abstractions should be used. Similar meaning stated in [14] that the current practice of MAS design tends to be limited to individual agents and small face-to-face groups of agents that operate as closed systems. MAS can be organized in variety of forms such as Hierarchy, Flat, Subsumption, and Modular organization. Not only that but also hybrids of these and others in addition to dynamic changes from one organization style to another are also possible [12][15].

Traditionally, MAS organization was doing completely in design time and system structure be fixed from the start to the termination of the system as shown in Fig.1. Furthermore, those systems concerned intra-group interactions (interactions inside one group of agents) with little attention to inter-group interactions (interactions between groups of agents). Also, in some of them the organization abstraction is not explicit and the responsibility of dynamic reorganization is given to individual agents in addition to their functional responsibilities the situation which called by Weyns [16] as dual responsibility which is very complex to engineer and not suitable for handling our real world complexity and other emerged characteristics such as highly distribution, unpredictability, uncertainty, continuous change and evolution. Those new characteristics should be reflected in current systems engineering methods.

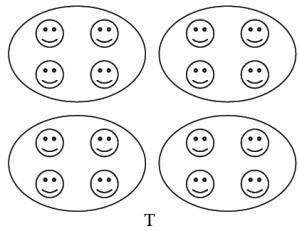

Fig.1 Static design of agents groups in design time

Currently, new engineering concepts imposed themselves on modern real-life systems, examples of these new concepts are dynamic reorganization, self-organization, and emergence. Dynamic reorganization can be defined as the change of MAS structure and behavior as a result of internal (local) or external (supervisory) demand. The external demand can be for example human intervention. The internal demand emerges from the system itself as an autonomous system to adapt to environments changes. Self-organization is a special type of dynamic reorganization; it is a dynamic higher level reorganization emerges as a result of internal system demand to adapt for dynamic environment changes. One of the valuable papers which addressed dynamic reorganization and self-organization in MAS is that of Picard et al. [17], who presented a comprehensive view of the organizational aspects in MAS from both agent-centered and organization-centered points of view. Fig.2 demonstrates the possible types of agent-organization relationship, as shown in Fig.2.a an individual agent searches for a suitable agents' organization and asked to join it to achieve its own goals. But in Fig.2.b the situation is reversed, an organization searches for an agent to give it a role to play inside it and the result will be achieving the required organization goals. The later approach is better because normally the organization has a more global view of the system than an individual agent.

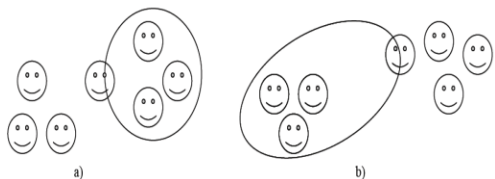

Fig.2 Types of agent-organization relationships, a) an agent search for organization to join, b) an organization search for an agent to employ.

The proposed NOSHAPE organizational model adopts the later technique shown in Fig.2.b. From the engineering perspective organizational models is the way through which the developer can design and plan how the MAS can be organized in design time and dynamically reorganize in run-time.

### III. THE NOSHAPE ORGANIZATIONAL MODEL

The main concern of organizational models is to describe the structural and dynamical aspects of organizations [18][30]. They provide a framework to manage and engineer organizations, dynamic reorganization, self-organization, emergence, and autonomy within multi-agent systems. Moreover, the underlying organizational model is responsible of how efficiently and effectively organizations carry out their assigned functional tasks, they have been recently used in agent theory for modeling coordination in open systems and to ensure social order in multi-agent system applications [19]. This section presents the proposed NOSHAPE organizational model for developing adaptive large-scale agent-based SCADA systems.

*A. General Description*

The philosophy of NOSHAPE is to consider an organization of agents as an explicit and static entity that has dynamic structural behaviors. In other words, the organization is something tangible and has a personality, it starts when the system starts and continues to exist until the system terminates. Moreover, new organizations can dynamically join the system in run-time. A noshapian MAS is a collection of organizations (an organization can be seen as a group of agents) able to overlap with each other to share or exchange roles/agents. As an example, Fig.3 shows a noshapian application world comprises four organizations each contains a number of individual agents. The system starts in the initial time ($T_i$) with no inter-organization relationships but after time goes and at $T_{i+k}$ the system structure changes and overlapping relationships are established among system organizations as a result of each organization lower level interactions and/or environment changes.

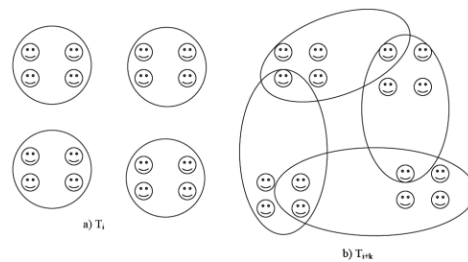

Fig.3 overlapping of organizations of agents within a noshapian MAS

The main principle behind the NOSHAPE model is that pairs of agents are more likely to be interacted if they are both members of the same organization(s), and less likely to be interacted if they do not share organizations. In other words, two agents can only interact if they belong to the same organization.

*B. Meta-Model*

Fig.4 provides the meta-model of a noshapian MAS. As shown in the figure, a noshapian MAS comprises a number of organizations; each of them contains one static role for organization structure management and many dynamic roles for application domain functional activities. Also, each organization is able to execute a number of structural dynamic behaviors by its static role relative to other organizations. The important dynamic behavior is to overlap with other organizations to share their dynamic roles. Initially, each organization is assigned a certain functional activity, which requires some environment resources, which are accessed by (or assigned to) the organization dynamic roles/agents. The lower level dynamic roles fire triggers to the organization static role declaring that they require a joint activity with other dynamic agents in other organizations to be able to achieve their assigned functional tasks. The static role responds to dynamic roles triggers and executes structural dynamic behaviors with other organizations, which encapsulate the required dynamic roles to share them.

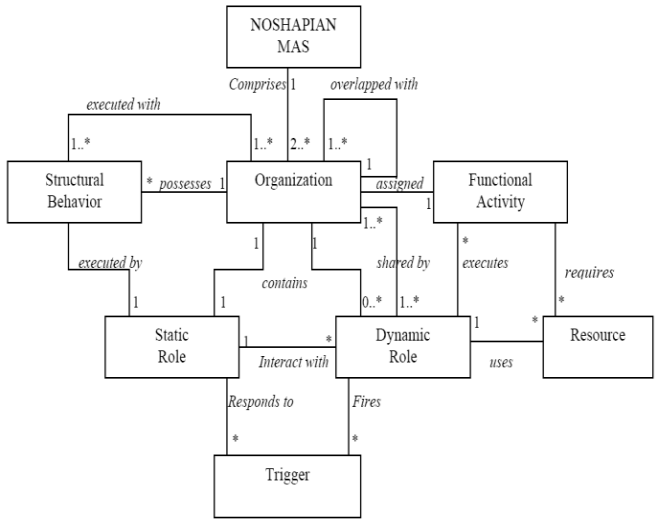

Fig.4 The Meta-Model of a one-world noshapian MAS

Fig.5 provides the anatomy of a noshapian organization. The figure illustrates the organization contents gradually from higher abstraction level (left) to the lower one (right). As shown in the figure, the organization is divided into two parts, static part which contains static roles which are responsible of the management of the organization structure. An example of a static role is the global supervisor (GS) which can be considered as the organization structure controller; it concerns the inter-organization interactions and manages the overlap and other dynamic behaviors with the other world organizations. The second part of a noshapian organization is the dynamic part which contains dynamic roles which are responsible of the application domain functional activities.

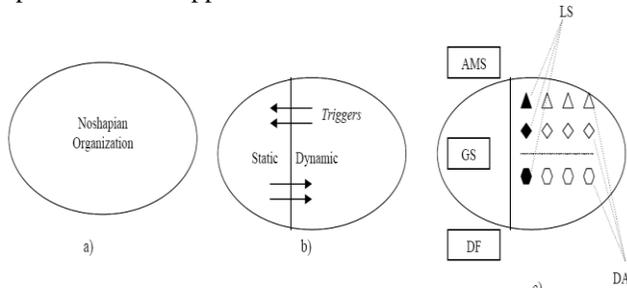

Fig.5 The anatomy of a noshapian organization

Dynamic roles can have many types (depending on the application domain); to free the global supervisor from caring directly about functional dynamic agents it's possible to assign a local supervisor (LS) to each type of dynamic roles. The responsibilities of the local supervisor are: (1) creates the required dynamic agents and configures them, (2) monitors dynamic agents health/performance and applies dynamic load balancing algorithms on dynamic agents, (3) recreates dead dynamic agents and reconfigure them, (4) receives triggers from dynamic agents and bypass them to the global supervisor, (5) assigns new added environment resources to dynamic agents. Fig.5.a indicates that the organization concept in the NOSHAPE model is a first class concept like the concept of agent. Fig.5.b shows the two parts of the organization and illustrates the interaction between them through triggers such as Trigger Required Joint Activity (TRJA), Trigger Finished Joint Activity (TFJA) and so on. Fig.5.c shows the internal structure of a noshapian organization, as shown in the figure the static part of the organization contains one static role (GS) or the global supervisor, and the organization dynamic part contains many dynamic roles/agents (DAs) with different types and each type is supervised by a local supervisor (LS). The figure shows two other components, the agent management system (AMS) and the directory facilitator (DF). Each organization should contain these two important services. The AMS has the responsibility of creating agents, removing agents, naming of agents etc, it provides the white page service to the organization. The DF provides the yellow page service inside an organization where the system agents advertise their services to enable other agents to know them. Any FIPA (Foundation for intelligent and physical agents)-compliant [20] agent development platform contains these two services such as JADE [21] which implements these services as agents. Thus when developing noshapian MAS applications it is assumed that these two services are available and can be used directly. That is why these two services are not included in the NOSHAPE meta-model shown in Fig.4. Fig.6 presents the adaptation control loop executed by the global supervisor of each noshapian organization.

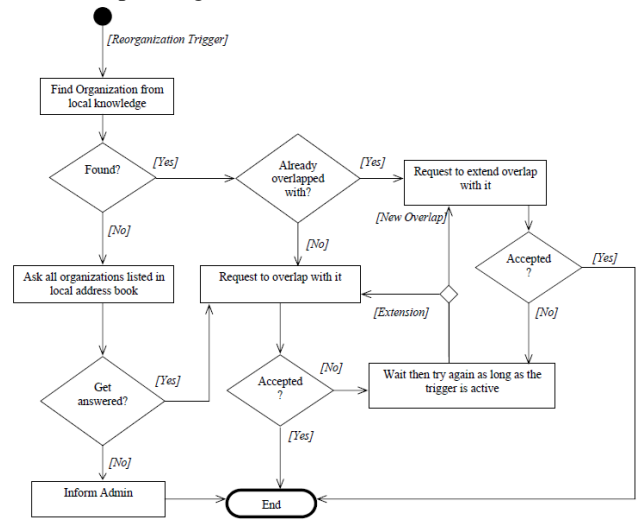

Fig.6 Adaptation control loop

It is required to emphasize here that the dynamic reorganization is excited by the organization internal functional activities (dynamic roles activities) in a bottom-up way as shown in Fig.7, this happens endogenously without external intervention and this complies with the requirement for self-organization. When there is a required joint functional activity with another organization a trigger fired from the organization dynamic part to its static part which in turn starts the dynamic reorganization process through the overlapping with other organizations to share the required dynamic roles/agents. After the overlapping had been done, the joint functional activity between the two organizations dynamic agents is pursued.

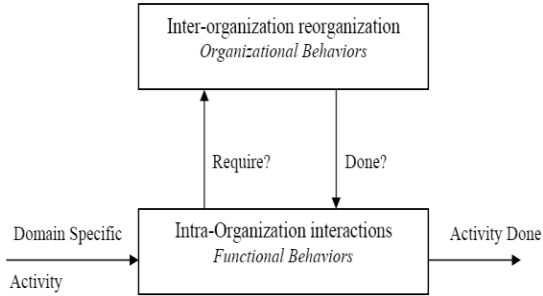

Fig.7 The dynamic reorganization excited by low level functional activity

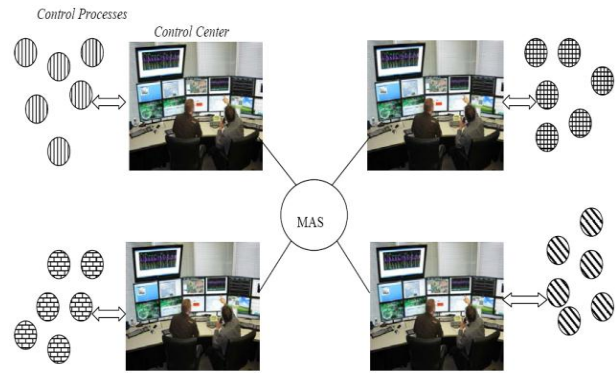

Fig.8 A large-scale distributed control center comprises four medium-scale control centers

## IV. THE PROPOSED ADAPTIVE SCADA

Two emerged concepts are definitely related to modern SCADA systems. The first one is the concept of Internet of Things (IoT) [22], which can be defined as the interconnection of uniquely identifiable physical embedded computing devices within the existing Internet infrastructure, is getting great attention and it is expected that in the near future most of nations' critical infrastructure utilities, and industrial activities will be connected globally using the Internet as the underlying network. The second concept is what is called system of systems (SoS) [23], which are large-scale concurrent and distributed systems the components of which are complex systems themselves. SoS are not only complex and large-scale but also they are characterized by decentralized, distributed, networked compositions of heterogeneous and autonomous components. Modern and future SCADA systems are considered as SoS because their increasing scale and complexity have reached a point that imposes qualitatively new demands on them.

Modern SCADA systems are an example of complex large-scale industrial networks. It concerns local/remote real-time control, supervisory, and monitoring of industrial processes, it is currently an interesting research area for both the industrial and academic communities specially when integrated with Internet forming what is called web-based SCADA [24][25]. SCADA systems had been evolved from being local small-scale to become global large-scale systems and in the near future it is expected that these systems evolve more and more to become ultra-large-scale systems. The case study presented here is concerned with the engineering of a large-scale distributed SCADA system comprises a number of medium-scale control centers (subsystems) located in different locations and each one is assigned a number of control processes or plants to supervise and control. In the following subsections we show how these complex systems can be engineered with MAS based on the proposed NOSHAPE organizational model for realizing dynamic reorganization with MAS.

### A. The Proposed Adaptive SCADA Architecture

The physical architecture of the system-to-be is shown in Fig.8; this system is considered as a large-scale distributed control center comprises a number of medium-scale control centers (initially 4).

Each control center controls, supervises, and monitors large number of control processes geographically distributed within the control center surrounding local environment. It is required that not only the operators in one control center can monitor and supervise their local control processes but also they are able to supervise, and monitor any control process located in the environment of another control center. This system is characterized to be open, highly distributed, and complex as the number of its subsystems can dynamically evolve horizontally (highly distribution) and vertically (increase scale and complexity of system information layers). Dynamically reorganized MAS are the best choice for engineering this type of systems because they are the most representative among artificial systems dealing with complexity and distribution [3].

Each control center will be modeled as a noshapian organization assigned a number of control processes controlled by programmable logic controllers (PLC). The PLCs can be considered as the environment resources of the control center. For the sake of simplicity and because our intention is to provide a prototype system based on the new proposed organizational model as a proof of concept, the number of system organization is supposed to be four and the number of control process assigned to each organization to be six, but with very large-scale systems these two numbers can be very large. The functionality of the system-to-be can be captured using the use cases artifact. A use case describes a required functional scenario in the system. The system higher level use cases are shown in Fig.9. There are only two real actors in the system-to-be, the operators and the control processes or industrial plants. The main use case of the system concerns providing the remote operator with an access to the required control processes. If the required control process belongs to the same organization of the remote operator agent, then no inter-organization dynamic reorganization is required. The remote operator agent is only needs to use the local yellow page service to find the provider (dynamic agent) of the required service or control process. But in case the required control process is located in a far environment and is under the direct supervision of another organization (control center) then an inter-organization dynamic reorganization is required, the organization hosts the remote operator agent will try to interact with the organization which has the required control process to share it through the overlap dynamic behavior between the two organizations. If the overlap process is successful, the remote operator should be notified and given the required information to be able to access the required control process and pursue the supervisory and control activities.

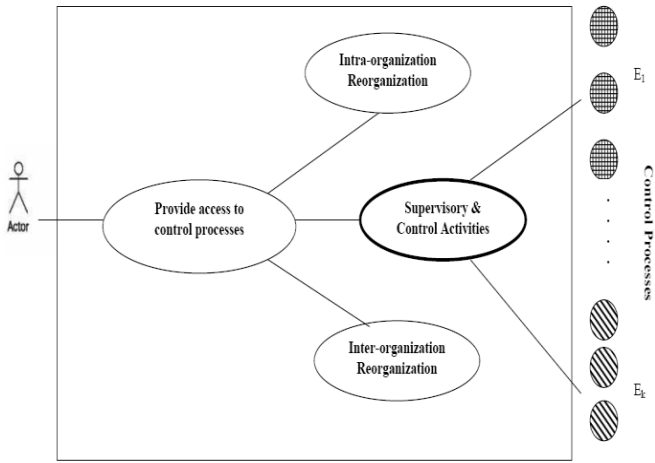

Fig.9 The higher level Use case of the proposed SCADA

The supervisory and control activities use case (shown in **bold** border in Fig.9) is a composite one comprises many other use-cases:

1. A use case for providing real-time monitoring to the operator.
2. A use case for receiving operator setpoints.
3. A composite one for providing higher level control such as, checking operator setpoints validity, forwarding operator valid setpoints to control processes, notifying operators with changed process data, providing global synchronization, and providing higher level control algorithms.

As we mentioned in the last paragraph of Section 1, the proposed adaptive SCADA system comprises two architectures. The internal one which is the architecture for modeling the application domain, and the external one which provides the adaptation layer based on proposed organizational model designed for MAS dynamic reorganization. The internal functional architecture was previously presented as a case study in [29]. This paper concerns only the adaptation external architecture with little focus on the functional one.

Table 1 presents the agents' types inside each organization according the NOSHAPE model specifications and suggests the responsibilities of each agent type. The responsibilities of each agent type divided into two types, interaction protocols for interactions with other agents if required and internal activity for executing the agent assigned functional tasks. We assume that the agents' responsibilities shown in Table I are obvious and self-explained and therefore there is no need to explain them in details for the sake of paper size.

TABLE I
RESPONSIBILITIES OF EACH AGENT TYPE INSIDE EACH ORGANIZATION

| Agent | Responsibilities |
|---|---|
| Global Supervisor GS | 1. Read configuration files<br>2. Create local supervisory agents (LS)<br>3. Configure local supervisory agents<br>4. Listening to local supervisory agents<br>5. Monitoring local supervisory agents<br>6. Manage organization structure<br>7. Search for required remote services<br>8. Dynamically Interact with other organizations |
| Local Supervisor LS | 1. Receive configuration from global supervisor<br>2. Creating dynamic agents<br>3. Configuring dynamic agents<br>4. Register services to local directory facilitator (DF)<br>5. Listening to dynamic agents<br>6. Monitoring dynamic agents<br>7. Recreating and reconfiguring dead dynamic agents<br>8. Executing load balance algorithms on dynamic agents<br>9. Receiving remote agents requests |
| Control Agent CA | 1. Receiving configuration from local supervisory<br>2. Register services to local directory facilitator (DF)<br>3. Connecting to its assigned control processes<br>4. Respond to remote operator agents<br>5. Provide higher level control algorithms<br>6. Send notifications to local supervisor<br>7. Respond to local supervisor |
| Remote Operator Agent RA | 1. Searching and subscribing to local DF for dynamic agents<br>2. Receiving notifications from local DF<br>3. Interacting with dynamic agents<br>4. Notifying LS if the required services are not available<br>5. Present process data to operators (textual, tabular, graphical,…etc)<br>6. Send operators set points to dynamic agents |

*B. The Proposed Adaptive SCADA Design*

The internal structure of each organization is shown in Fig.10 where $P_i$ represents the control processes; $C_i$ (control agents) and $R_i$ (remote operator agents) represent the organization dynamic agents. LS is the local supervisor, GS is the organization global supervisor, and AMS and DF are the agent management and directory facilitator respectively. Initially and just after starting up the system-to-be only the GS created then it reads the system configuration from an XML file includes the control processes configurations such as process name, connection interface and the process variables required to be monitored.

Another configuration XML-file is provided to the GS includes the initial information about its acquaintance organizations such as their names, addressees etc. After reading the configuration XML-files the GS processes these configurations it creates local supervisors (this happens by the interaction with the organization agent management system or AMS service) then it sends the configuration data to the created LS through a request message. The LS processes the configuration data and then creates the dynamic agents which will connect to the control processes using OPC protocol [27] for the sake of control systems interoperability. Then LS assigns to each dynamic agent its control process to supervise, in this prototype we assume one control process per each dynamic agent but it is possible to assign more than one control process to a dynamic agent. After that, each dynamic control agent registers its assigned service to the local DF to enable other dynamic agents to find it.

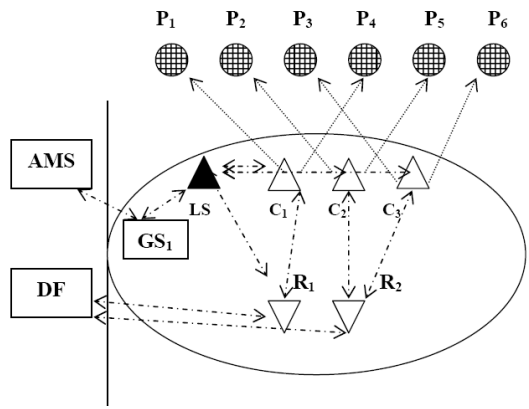

Fig.10 The architecture of a noshapian Organization and its agents' local interactions and connections, it represents a medium control center

It has to be pointed out here that there are two types of dynamic agents in this prototype, control agents (Cs) for direct connection to control processes and remote operator agents (Rs) for remote operator access to the system. A remote operator agent ($R_i$) after starting searches for a certain service through the organization's local DF which then delivers the required service provider ($C_i$) to it if available. The dynamic control agents are created by the LS but the dynamic remote operator agents are created manually by the operators as required, however, both of them are supervised by the LS. To demonstrate the nature of inter-organization interactions, Fig.11 provides a possible scenario of system evolution from the initial start up of the system at T0 to time T7. The Contract Net interaction protocol [28] is used for coordination between agents' organizations.

*C. The Proposed Adaptive SCADA Implementation*

The implementation phase is concerned with moving the automated system-to-be from the development status to the production status. It implicitly includes the deployment of the new system to its target working environment. The JADE framework [21][26] was chosen to implement the proposed adaptive large-scale agent-based SCADA system. JADE is a software Framework fully implemented in the Java language. It simplifies the implementation of multi-agent systems through a middle-ware that complies with the FIPA specifications and through a set of graphical tools that support the debugging and deployment phases. Table II provides a possible mapping from NOSHAPE concepts to JADE constructs. The proposed adaptive SCADA can be implemented as independent JADE platforms interact together through a wide area network such as the Internet. JADE as a middle-ware framework provides all the required low level services to enable flexible agents' interactions [31].

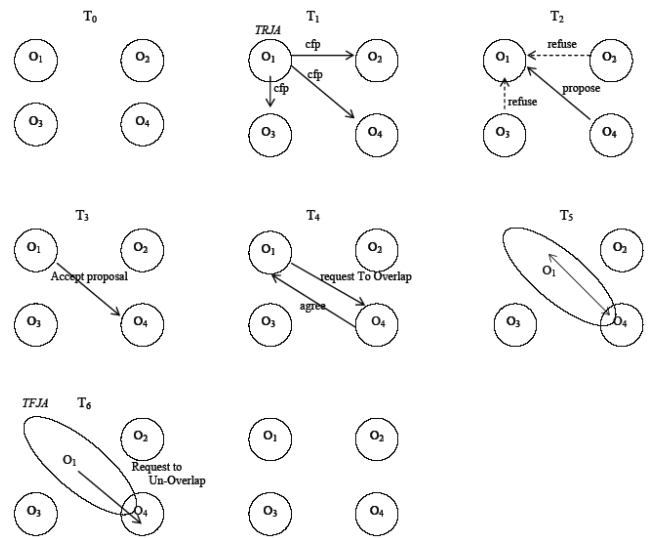

Fig.11 An example of inter-organization dynamic reorganization

TABLE II
MAPPING NOSHAPE CONCEPTS TO JADE CONSTRUCTS AND SERVICES

| NOSHAPE Concept | JADE Construct |
|---|---|
| Noshapian Organization | JADE Platform |
| White page service | JADE AMS |
| Yellow page service | JADE DF |
| Static Role | JADE Agent |
| Dynamic Role | JADE Agent |
| Finite State Machine (FSM) | JADE FSMBehaviour |
| Dynamic Structural Behaviors | JADE Interaction Protocols |
| Overlap meaning | Registering to a remote DF |
| Functional Activities | JADE Agent Behaviors |
| NOSHAPE species and Their relations | JADE Ontology Support |

Each JADE platform can be distributed across many machines. It is intuitively obvious that each noshapian organization in the system-to-be can be modeled as an independent JADE platform and each organization white page and yellow page services can be matched to the JADE platform AMS and DF services respectively. Each control center in the proposed SCADA system is modeled as a noshapian organization contains two types of agents' roles. Static roles (i.e., the organization global supervisor (GS)) are responsible of the management of the organization structural behaviors and the local supervisor (LS) for managing the local dynamic agents. The dynamic roles are responsible of the domain specific functional activity inside each organization and can be shared among organizations. In this case study two types of dynamic agents are used. The first is the control agents (Cs) which can be considered as the service providers and the second type is remote operator agents (Rs) which is the operator interface to the system to enable the operator to get access to control processes. The internal behavior and interaction behaviors of each agent were specified and implemented in JADE according to the responsibilities of each agent as presented in Table I. Not all system agents need to be provided by a graphical user interface (GUI) except the remote operator agent which is the operator interface point to the system and he uses it to supervise and control physical control processes. A simple GUI for remote operator agents is shown in Fig.12.

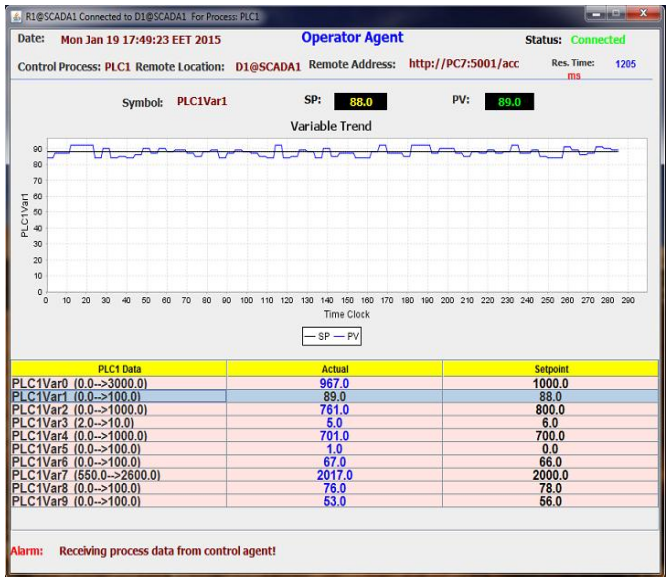

Fig.12: The human machine interface (HMI) designed for a remote operator agent

### D. The Global Overview and performance Evaluation

Typically it is difficult for academic researchers to test large-scale systems with physical work environments, and usually they have to create simulation environments. Therefore, to test the developed adaptive agent based SCADA we created a simulation environment based on randomly generated OPC process data. System organizations are deployed as JADE platforms geographically separated and communicate through large local area network (LAN). The global overview of the system is shown in Fig.13. Each JADE platform deployed on a sub-LAN comprises a set of hosts. The adaptive SCADA enabled us to run operator agents on any host to access any control process in the system.

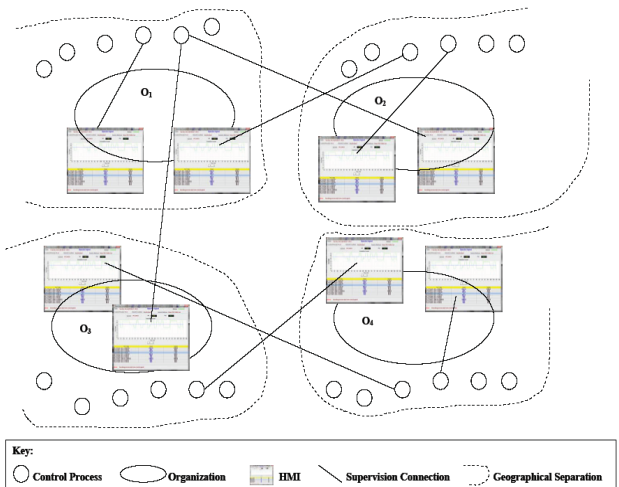

Fig.13 System global overview

To demonstrate the effect of the realized system adaptivity through dynamic reorganization on system performance, we carried out an experiment whose results are presented in Table III.

TABLE III
PERFORMANCE EVALUATION SHOWS THE EFFECT OF DYNAMIC REORGANIZATION

| $O_1$ | | $O_2$ | | $O_3$ | | $O_4$ | |
|---|---|---|---|---|---|---|---|
| $O_i.PLC$ # | $T_s$ | $O_i.PLC$ # | $T_s$ | $O_i.PLC$ # | $T_s$ | $O_i.PLC$ # | $T_s$ |
| $O_1.1$ | 0.8 | $O_2.7$ | 1.1 | $O_3.13$ | 1.3 | $O_4.19$ | 1.3 |
| $O_2.7$ | 3.6 | $O_1.1$ | 3.8 | $O_3.14$ | 0.8 | $O_4.20$ | 0.8 |
| $O_2.8$ | 2.8 | $O_1.2$ | 2.7 | $O_3.15$ | 0.9 | $O_4.21$ | 0.8 |
| $O_2.8$ | 0.8 | $O_1.3$ | 2.4 | $O_1.1$ | 3.8 | $O_4.22$ | 0.6 |
| $O_3.14$ | 3.8 | $O_3.13$ | 4.1 | $O_1.5$ | 2.7 | $O_4.23$ | 1.2 |
| $O_3.15$ | 2.7 | $O_3.14$ | 2.7 | $O_2.7$ | 3.7 | $O_4.24$ | 1.1 |
| $O_4.19$ | 3.9 | $O_4.19$ | 3.8 | $O_2.8$ | 2.6 | $O_1.1$ | 3.9 |
| $O_4.20$ | 2.5 | $O_4.20$ | 2.8 | $O_4.19$ | 3.9 | $O_1.4$ | 2.8 |
| $O_4.21$ | 2.6 | $O_4.21$ | 2.6 | $O_4.20$ | 2.6 | $O_3.13$ | 3.7 |
| $O_3.17$ | 2.6 | $O_2.8$ | 0.9 | $O_3.16$ | 0.8 | $O_3.14$ | 2.6 |
| $O_4.23$ | 2.6 | $O_1.5$ | 2.4 | $O_3.17$ | 1.1 | $O_2.7$ | 3.8 |
| $O_4.20$ | 0.8 | $O_1.6$ | 2.7 | $O_3.18$ | 0.8 | $O_2.8$ | 2.7 |

Table III is horizontally divided according to the initial system organizations $\{O_1, O_2, O_3, O_4\}$. The vertical dimension of the table represents the order of launching a remote operator agent in each organization searching for a certain PLC where $O_i.j$ means the launching of a remote operator agent inside $O_k$ where "j" represents the required PLC number belongs to $\{O_i\}$. The content of the table represents the time in seconds ($T_s$) required for a remote operator agent to wait until it has access to the required service (PLC). This time represents the interval taken by the dynamic reorganization process of the system to make the required service available to the concerned remote operator agent. Remember that each organization contains two types of dynamic agents, control agents for interfacing the system with control processes, and operator agents which provide the operator with the human machine interface (HMI) control and supervision of control processes. And also remember that in each organization an operator can launch an operator agent to have access to any PLC interfaced by any dynamic agent located in any organization.

Let us analyze the results related to organization $O_1$ shown in first main column (from the left), as shown, firstly an operator launched an operator agent to get access to PLC1 which belongs to a control agent situated in the same organization $\{O_1\}$ and the response time to get access to PLC1 is just one 0.8 seconds. After that, another operator launched another HMI agent to get access to PLC7 which belongs to organization $\{O_2\}$ so the response time is 3.6 seconds (but now an overlap relationship established between $\{O_1\}$ and $\{O_2\}$), the effect of this overlap can be seen in the next row of the table where an operator launched an HMI agent to access PLC8 which belongs to $\{O_2\}$ but $\{O_1\}$ is already overlapped with $\{O_2\}$ so it just need to ask $\{O_2\}$ to add the control agent for PLC8 to the shared agents between them and that decreases the response time to be 2.8 seconds. The fourth table row provides an interesting result, an operator launched an HMI agent in $\{O_1\}$ to get access to PLC8 in $\{O_2\}$ and he got a response time equals 0.8 seconds because the control agent which has interface with PLC8 in $\{O_2\}$ has already registered itself in $\{O_1\}$ local DF and that means there is no need to dynamic reorganization because $\{O_1\}$ is currently overlapped with $\{O_2\}$.

### V. CONCLUSIONS AND FUTURE WORK

The increasing complexity, heterogeneity, and openness of modern SCADA systems have reached a point that imposes new demands on their engineering technologies. Conventional engineering approaches, methods, and technologies will stand powerless in front of future SCADA increase in scale and complexity either vertically. Building adaptive SCADA able to adapt fluidly changing working environments is currently an active research area. Multi-agent systems emerged as a new engineering style for building adaptive large-scale systems. They model the system as a distributed autonomous agents cooperated together to achieve the global system goals. The ability of agents to dynamically reorganize to adapt environments changes is a key feature provided by multi-agent systems. This research proposes a MAS-based approach for developing an adaptive large-scale SCADA system consists of many medium-scale control centers geographically distributed but interact together to provide real-time supervision and control of a large number of production processes. System adaptivity is realized thanks to the proposed organizational model which enabled the system-to-be to dynamically reorganize to adapt environment changes. This type of systems are getting more complex, open, and critical because they are adopted to remotely supervise and control most critical infrastructure utilities such as power grids, water transportation. As a future work it is still required to use the proposed adaptation architecture for building very large-scale SCADA systems in addition to using proper security methods and techniques.